\documentclass[journal]{IEEEtran}
\usepackage{cite}

\usepackage[cp1252]{inputenc} 
\usepackage{textcomp}
\usepackage{amsmath}
\usepackage{amssymb}
\usepackage{graphicx}
\usepackage{esint}
\usepackage{sidecap}
\usepackage{float}
\usepackage{color}

\begin{document}
\title{The role of the Fermi level pinning in gate tunable graphene-semiconductor junctions}
\author{Ferney A. Chaves, David Jim\'enez

\thanks{The authors are with the Departament d\textquoteright{}Enginyeria Electr\`{o}nica, Escola d\textquoteright{}Enginyeria, Universitat Aut\`{o}noma de Barcelona, Campus UAB, 08193 Bellaterra, Spain (e-mail: ferneyalveiro.chaves@uab.cat; david.jimenez@uab.cat).}

}
\maketitle

\begin{abstract}

Graphene based transistors relying on a conventional structure cannot switch properly because of the absence of an energy gap in graphene. To overcome this limitation, a barristor device was proposed, whose operation is based on the modulation of the graphene-semiconductor (GS) Schottky barrier by means of a top gate, and demonstrating an ON-OFF current ratio up to $10^5$. Such a large number is likely due to the realization of an ultra clean interface with virtually no interface trapped charge. However, it is indeed technologically relevant to know the impact that the interface trapped charges might have on the barristor's electrical properties. We have developed a physics based model of the gate tunable GS heterostructure where non-idealities such as Fermi Level Pinning (FLP) and a ``bias dependent barrier lowering effect" has been considered. Using the model we have made a comprehensive study of the barristor's expected digital performance.  
\end{abstract}


\section{Introduction}
Graphene is one of the most studied materials because of its unique properties related to its two dimensional nature. It offers the possibility of integration with the existing semiconductor technology for next-generation electronic and sensing devices. In particular, its high conductivity makes it suitable for replacing traditional metal electrodes in Schottky diodes \cite{Tongay, Parui, Yim, Chen, Sinha}. The graphene-semiconductor (GS) Schottky diode structure has been a platform to recent studies in interface transport mechanisms as well as for applications in photodetection, high-speed communications, solar cells, chemical and biological sensing, etc. \cite{Kopens, Lis, Li, Lancellotti, Miao, Kim}. However, despite the intensive researches into graphene electronics, graphene transistors exhibit a very poor ON-OFF current ratio ($I_{on}/I_{off}$), insufficient for digital applications, being the absence of an energy gap in graphene the reason behind. Few years ago, H. Yang et al. \cite{Samsung} proposed a three terminal device termed as "Barristor" to help overcoming this limitation, demonstrating an impressive $10^5$ ON-OFF current ratio. In the barristor a top gate is added to the GS junction to control the Schottky barrier height (SBH) and so to achieve a large modulation of the diode current. In Yang's work an important aspect to suppress the formation of GS interface states and to avoid the appearance of Fermi-level pinning (FLP) was the optimized transfer process. In contrast, there are other examples where partial FLP plays an important role. For instance, Kim et al. \cite{Kimour} demonstrated a graphene/GaSe dual heterojunction device where a tunable current rectification was observed by the modulation of the Fermi level of graphene with the gate voltage. The tunability of the Femi level was slightly weakened because of partial FLP produced by interface states in the GaSe. In this context, a thorough understanding of the physics and the potentialities of the gate controlled GS diode is of great importance and it must be subject of systematic investigation.

\begin{figure}
\includegraphics[scale=0.48]{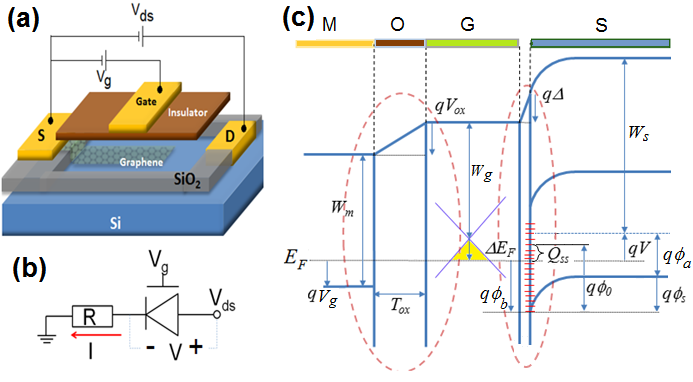}
\centering
\caption{(a) Sketch of the barristor device, (b) barristor's equivalent circuit, (c) band diagram of the MOGS heterostructure. There is an interface layer between G and S with thickness $d$. In order to be consistent with the sign of the charges in our model, in the figure $V_{ox}, \Delta, V_g, \Delta E_F$ and $\phi_s$ are positives and $V$ is negative.}
\label{device}
\end{figure}

\begin{figure*}[htb]
\centering
\includegraphics[width=1\textwidth,scale=0.14]{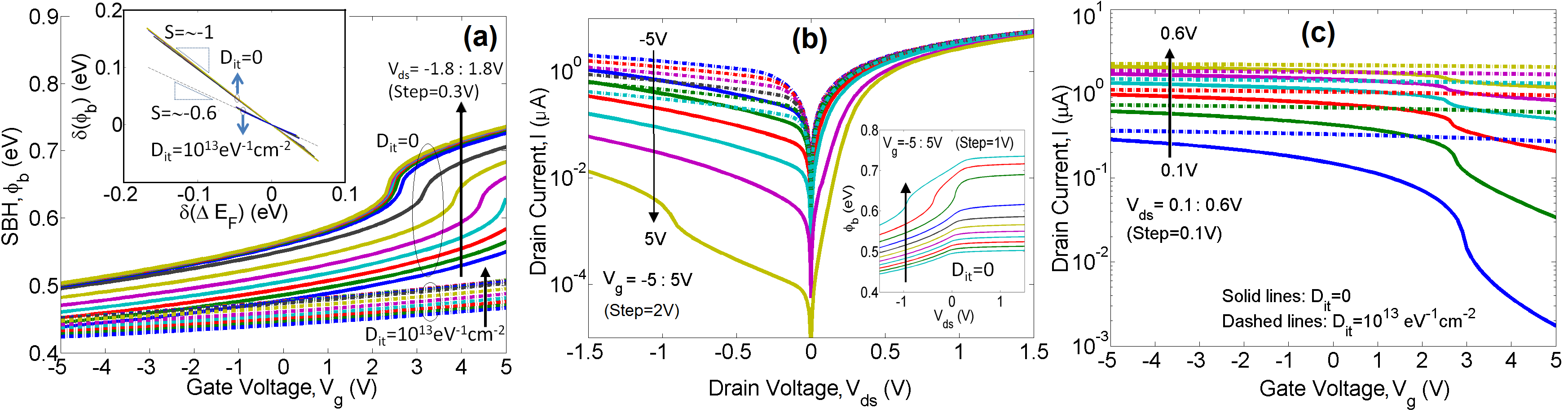}
\caption{Effect of the interface trapped charge on the SBH and current. Solid lines: without FLP and dashed lines: with FLP. (a) SBH curve for different values of $V_{ds}$. The inset shows $\delta\phi_b$ vs $\delta(\Delta E_F)$ for a wide range of $V_g$ (from -5V to 5V)  and $V_{ds}$. The dashed line in the inset is a extrapolation of the simulated data. (b) Barristor's output characteristics. The inset displays the bias dependent barrier lowering effect. (c) Barristor's transfer characteristics. In our simulations we have assumed a neutral level $q\phi_0=0.4$ eV, which is a typical value for silicon \cite{Sze}}. 
\label{panel1}
\end{figure*}

In this work, we extend the current understanding by proposing a physics based theoretical study of the electrostatics and $I-V$ characteristics of the barristor taking into account the effects of possible interface trapped charges, resulting in FLP. We also explore the impact of scaling the device through the reduction of the gate oxide thickness.

\section{Model}
Fig. \ref{device}a shows a sketch of the barristor studied in this paper. The bias voltage $V_{ds}$ produces a flow of carriers from source to drain forcing them to go from monolayer graphene to semiconductor, where a Schottky junction is formed. The SBH is modulated by a top gate voltage $V_g$, which produces a field-effect through an insulator of thickness $T_{ox}$. The equivalent circuit of the Barristor here considered is shown in Fig. \ref{device}b, where $R$ represents the series resistance, including both contact (source and drain) and channel (graphene and silicon) resistances, $V$ is the voltage drop across the Schottky juntion and $I$ the current flowing across the barristor. In order to get a better understanding of the electrostatics, Fig. \ref{device}c shows the band diagram of the Metal/Oxide/Graphene/Semiconductor (MOGS) vertical structure, where a p-type semiconductor has been assumed over here, without loss of generality. Here $W_m$, $W_g$, and $W_s$ are the gate metal, graphene and semiconductor work functions, respectively. $V_{ox}$ and $\Delta$  are the voltage drops across the gate oxide and the GS interface, respectively. $\Delta E_F$ is the electrostatically induced shift of the graphene Fermi level respect to the Dirac point, $q\phi_b$ is the value of the SBH, $\phi_s$ is the surface potential of the semiconductor and $q\phi_a$ is the difference between the Fermi level and the top of the valence band taken in the semiconductor's bulk. In our model we have assumed an interface layer of thickness $d=0.3$ nm \cite{Dang}. Aditionally, in order to take into account possible FLP because of surface states in the semiconductor, we have included a finite interface trapped charge $Q_{ss}$ in the model, assuming  that those states are filled according to the graphene Fermi level \cite{Gomila}. The current characteristics of the device have been computed following a Landauer transport theory for the thermionic emission considering the finite density of states of graphene $D=2\pi^{-1}(\hbar v_f)^{-2}\vert \Delta E_F \vert=D_0\vert \Delta E_F \vert$ ($v_f$ is the Fermi velocity in graphene and $\hbar$ the reduced Planck's constant) \cite{Sinha}:

\begin{equation}
I=I_0\left( \dfrac{\phi_b}{v_t}+1\right)e^{-\phi_b/v_t}\left[e^{V/\eta v_t}-1 \right],
\label{eq:current}  
\end{equation}

where $I_0=q^3v_t^2D_0A/\tau$, $v_t=k_BT/q$, $A$ is the effective area of the Schottky diode, $q$ is the elementary charge, $k_B$ the Boltzmann constant, $T$ the temperature, $\eta$ is the ideality factor and $\tau$ is the time scale for carrier injection from the contact.
\begin{subequations}
\label{eq:eltc}
\begin{align}
		 &Q_m+Q_g+Q_{s}+Q_{ss}=0\\
         &W_{g}+\Delta E_{F}=W_{m}+qV_{ox}-qV_g,\\
         &W_{g}+\Delta E_{F}+q\Delta=W_{s}-q\phi_s-qV\\
         &\phi_b=\phi_s+\phi_a+V    
\end{align}        
\end{subequations}
The SBH of the barristor is determined by solving Equations (2a)-(2d), which arise from the following conditions: (i) the total charge density in the heterojunction, including the gate contact metal charge $Q_m$, the graphene layer charge $Q_g$, the semiconductor charge $Q_s$, and a possible interface trapped charge on the semiconductor $Q_{ss}$ must be conserved (Equation (2a)) and (ii) the sum of voltage drops around any loop from the band diagram (see Figure \ref{device}c) should be equal to zero (Equations (2b)-(2d)). Also the following relations are satisfied: $Q_g=(qD_0/2)\Delta E_F\vert\Delta E_F\vert$ \cite{Castro}, $Q_{s}=-\sqrt{2q\epsilon_{s}N_A\phi_s}$ and  $Q_{ss}=-qD_{it}(q\phi_b-q\phi_0)$, where $q\phi_0$ is the neutral level (above $E_V$) of interface states \cite{Sze}.  The parameters $\epsilon_{s}$, $N_A$ and $D_{it}$ refer to the permittivity, doping concentration and interface trapped charge density of the semiconductor, respectively. The quantity in parentheses in $Q_{ss}$ is just the energy difference between the graphene Fermi level and the neutral level, so when they are the same, the net interface trapped charge is zero. In addition, the voltage drops across the oxide and interface layer are related with the charges as $V_{ox}=Q_m/C_{ox}$ and $\Delta = -(Q_{s}+Q_{ss})/C_d$, respectively. Here $C_{ox}=\epsilon_{ox}/T_{ox}$ and $C_d=\epsilon_d/d$ describe the gate and interface layer capacitances per unit area. Finally, the series resistance $R$ is related to the voltage drop across the Schottky juntion according to:
\begin{equation}
V=V_{ds}-IR.
\label{eq:VVds}
\end{equation}
By combining Eqs. 1-3 we can self-consistently solve both device's electrostatics and I-V characteristics (see Appendix A for an explanation). The main results revealing the impact of both FLP and scaling effects are shown in Figs. 2-5. To validate our model we have benchmarked it with experimental results from two kind of barristors operating in opposite limits (see Appendix D): (i) a barristor based on a p-type silicon substrate and $SiO_2$ as gate insulator \cite{Samsung} operating in the Schottky limit (no FLP) and (ii) a barrisor based on a GaSe substrate and $Al_2O_3$ as gate insulator \cite{Kimour} working in the Mott limit (strong FLP). We have assumed in our model the parameters reported in Table I, unless otherwise stated. 

\section{Results and Discussion}

Because the injection of the majority carriers (holes) from graphene to silicon is determined by $\phi_b$, the top gate modulates the magnitude of the current $I$. Fig. \ref{panel1}a shows, for two extreme cases, how the SBH can be modulated: i) without FLP, where $D_{it}=0$, and ii) with partial FLP, where $D_{it}$ has been assumed as $10^{13}$  eV$^{-1}$cm$^{-2}$. It is worth noting that due to the coupling among Eqs. 2a-2d our model predicts, in general, that the SBH not only depends on $V_g$ but also on $V_{ds}$, i.e. there is a ``bias dependent barrier lowering effect", similarly to the Drain Induced Barrier Lowering (DIBL) effect in short channel MOSFETs. In this sense, a barristor with p(n)-type semiconductor exhibits a reduction of its SBH when $V_{ds}$ negatively (positively) increases. 

From the inset of Fig. \ref{panel1}a we observe that there is a correlation between changes in the SBH and changes of the Fermi level shift, namely $\delta\phi_b\approx-\gamma	\delta(\Delta E_F)$. In the Schottky limit ($D_{it}=0$), $\gamma=1$ indicates that the Fermi level shift of graphene in absence of FLP is fully responsible for the variation of $\phi_b$. However, in a condition of partial FLP ($D_{it}\sim10^{13}$ eV$^{-1}$cm$^{-2}$) our simulations show $\gamma=0.6$, which is a clear indication of a loss of sensitivity of the SBH with $\Delta E_F$ and therefore with the gate voltage. An algebraic manipulation of Eq. 2c  (assuming $Q_{ss}>Q_s$), allows us to obtain the following analytical expression:

\begin{equation}
q\phi_b=\dfrac{q\phi_{b0}-\Delta E_F}{1+q^2D_{it}/C_d},
\label{eq:phi_b}
\end{equation}

where $q\phi_{b0}=\left( W_{sg}+q\phi_a+q^3D_{it}\phi_0/C_d\right)$ and $W_{sg}=W_s-W_g$. From Eq.  \ref{eq:phi_b} we can see the role played by both $D_{it}$ and $q\phi_0$ on the determination of SBH. The effects of the FLP on other electrical properties of the barristor will be shown below.

\begin{table}
\caption{Values of the parameters used in this work. The symbol ``*" means assumed value}
\resizebox{8.2cm}{!} {
\begin{tabular}{|c|c|c|c|c|c|c|c|}
\hline 
$W_m$(eV) & $W_g$(eV) & $W_{s}$(eV) & $\epsilon_{ox}(\epsilon_{0})$ & $\epsilon_{d}(\epsilon_{0})$ & $\epsilon_{s}(\epsilon_{0})$ & $R(k\Omega)*$ & $T(K)$ \\ 
\hline 
5.54 & 4.50 & 5.01 & 3.9 & 1 & 11.7 & 250 & 300 \\ 
\hline 
$v_f$(cm/s) & $d$(nm)* & $N_A$(cm$^{-3}$) & $E_g$(eV) & $q\phi_0$(eV)* & $\eta$ & $A$(cm$^{2}$)* & $\tau$(s)* \\ 
\hline 
$10^8$ & 0.3 & 5$\times10^{16}$ & 1.12 & 0.4 & 1.1 & $10^{-5}$ & $10^{-13}$ \\ 
\hline 
\end{tabular}
}
\label{tabla1}
\end{table}

\begin{figure}
\centering
\includegraphics[scale=0.17]{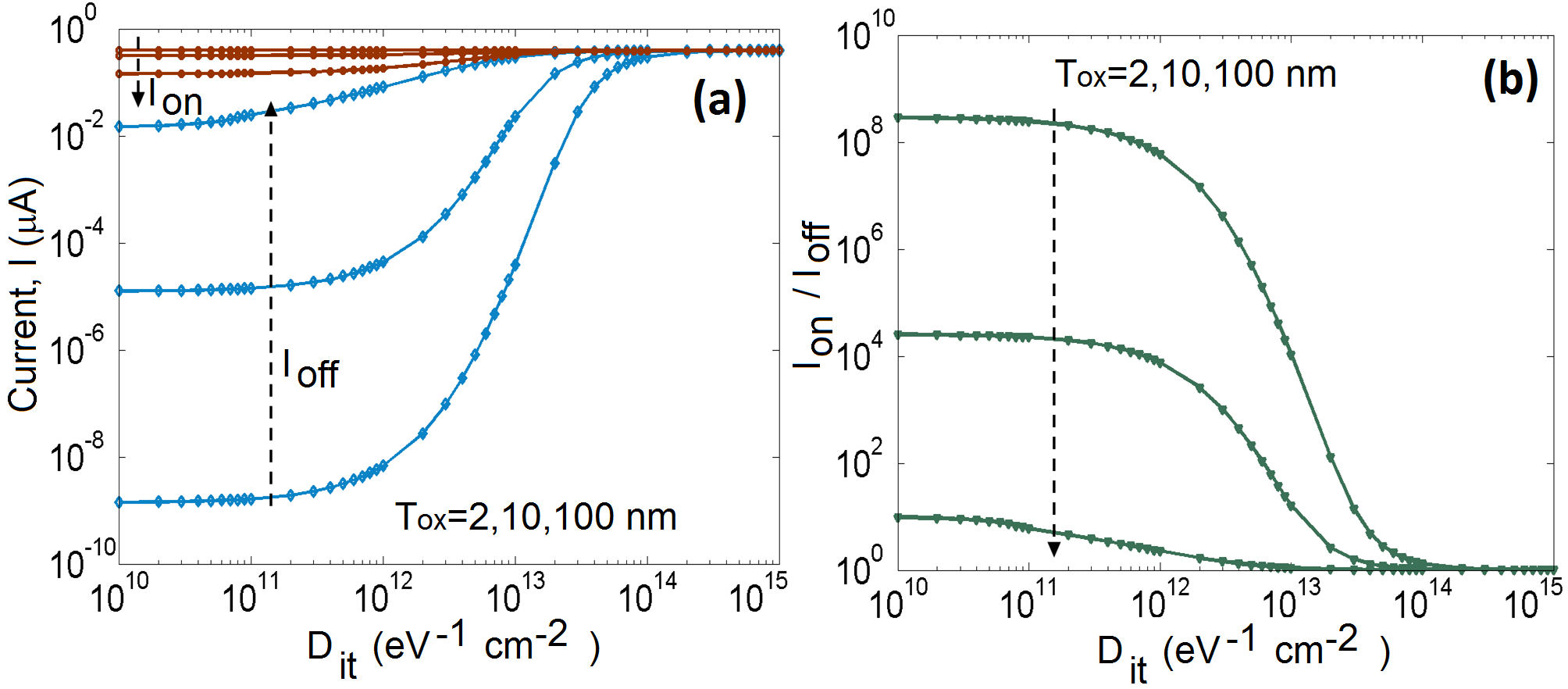}
\caption{(a) On current $I_{on}$ and off current $I_{off}$ as a function of $D_{it}$. (b) Ratio $I_{on}/I_{off}$ as a function of $D_{it}$. The dashed arrows indicate the increasing direction of the oxide thickness. In these curves $V_{g,on}=0.1$V, $V_{g,off}=3$V and $V_{ds}=0.1$V}
\label{Fig3}
\end{figure}

\begin{figure}
\centering
\includegraphics[scale=0.15]{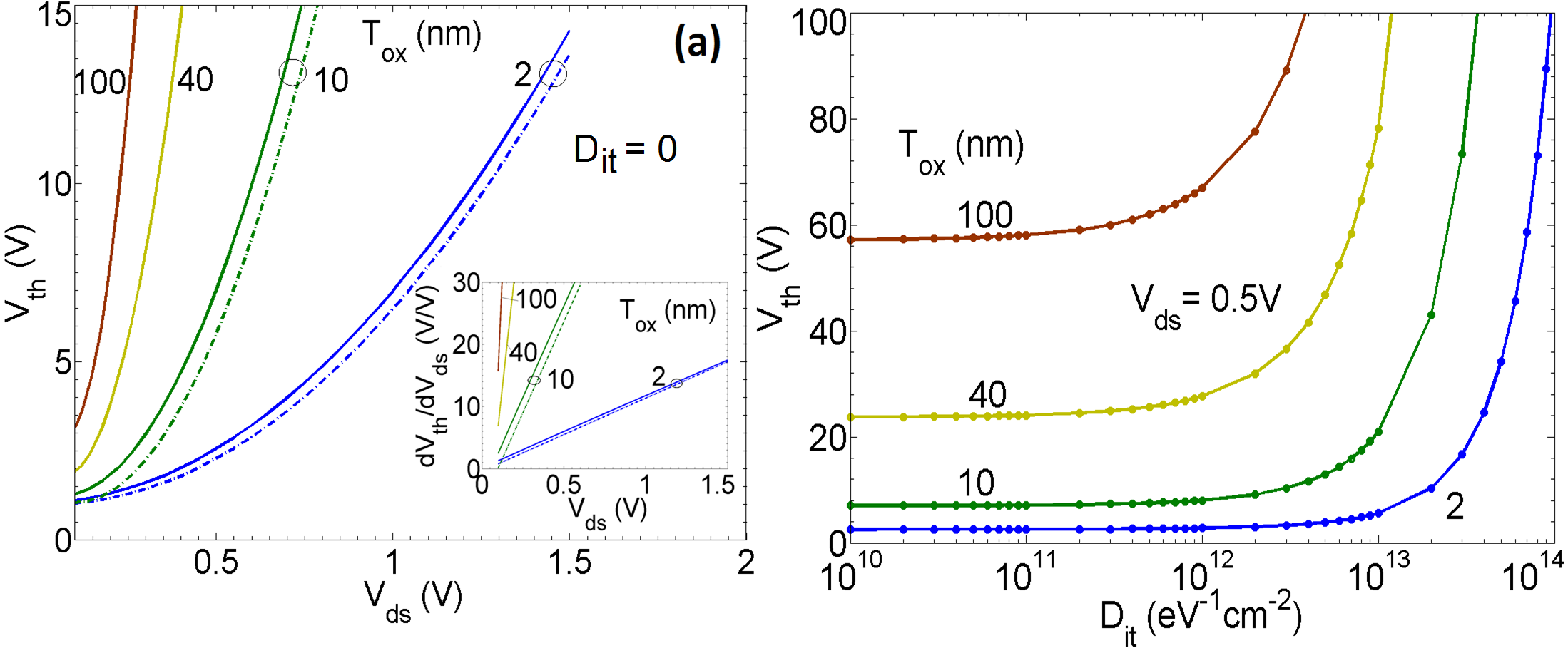}
\caption{(a) Threshold voltage as a function of $V_{ds}$. Solid lines: numerical results; and dashed lines: model given by Eq. \ref{eq:Vth}. Inset: the derivative of the threshold voltage with respect to $V_{ds}$. (b) Dependence of the threshold voltage on $D_{it}$. Here we have assumed $u_0\sim50$.}
\label{Fig4}
\end{figure}
Next, we analyze the output characteristics of the barristor (Fig. \ref{panel1}b). As for the case of no FLP, a strong rectification could be induced provided $V_g>>0V$. In contrast, if FLP comes into play, the SBH becomes almost insensitive to the gate voltage (Fig. \ref{panel1}a) and rectification fades out. Unlike typical FET-like device operation, the diode current does not saturate as $V_{ds}$ increases, but increases almost linearly. However, near the diode turn-on regime ($\sim$ 0-0.3V), $I$ varies by several orders of magnitude as $V_g$ changes, resulting in a switching operation with a large $I_{on}/I_{off}$ ratio (see Appendix B). The inset of Fig. \ref{panel1}b shows the dependence of SBH on $V_{ds}$ for several gate voltages. In this case, two regions with different behavior can be observed: (i) at $V_{ds}\lesssim 0$ there is a nearly linear dependence and (ii) at $V_{ds}\gtrsim 0$ the SBH saturates due to the effect of the series resistance. As $T_{ox}$ is reduced, the SBH becomes less sensitive to $V_{ds}$ (specially for negative gate voltages), but more sentitive to $V_g$ as shown in Appendix B.

\begin{figure*}[htb]
\centering
\includegraphics[width=1\textwidth,scale=0.14]{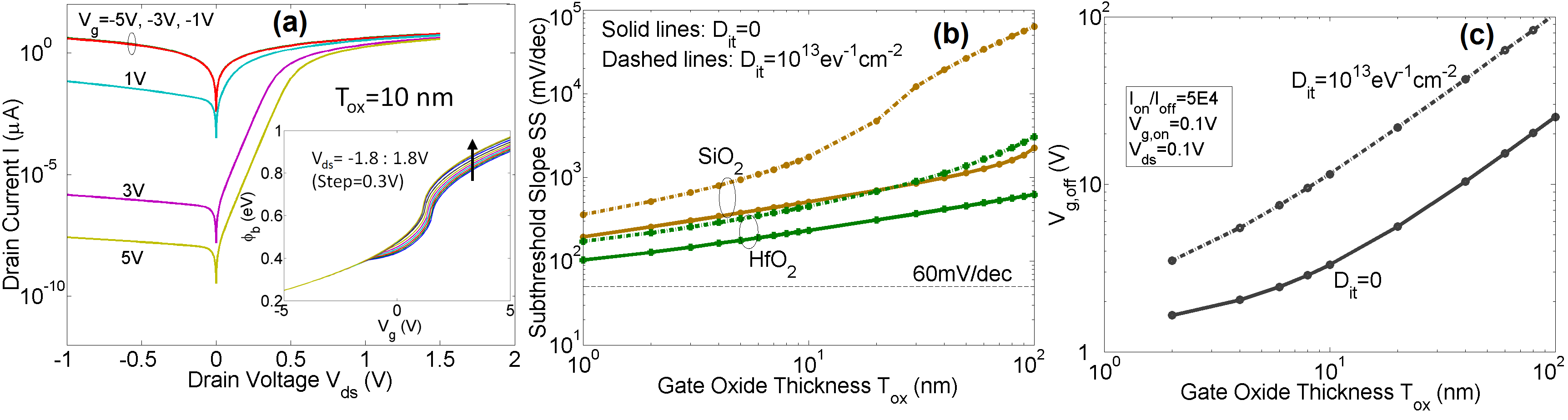}
\caption{Effect of the oxide thickness scaling on the electrical characteristics. (a) Output characteristics without FLP for $V_g$ varying from -5 V to 5 V. The inset shows the SBH as a function of the gate voltage for different $V_{ds}$. (b) Average subthreshold slope as a function of $T_{ox}$ in the range $V_g=1 - 3$ V for two different insulator materials. (c) $V_{g,off}$ as a function of the oxide thickness keeping $I_{on}/I_{off}$ and $V_{g,on}$ constants. In (b)-(c) solid (dashed) curves correspond to simulations without (with) FLP.}
\label{panel2}
\end{figure*}

Fig. \ref{panel1}c shows the transfer characteristics of the barristor for both no FLP and partial FLP cases. In the former case, the curves for $V_{ds}>0$ exhibit the greatest on current, and among them, the corresponding to low values of $V_{ds}$ have the best ON-OFF current ratio. If the on(off) state is defined at the bias point $V_g=0.1$ V ($V_g=3$ V) with $V_{ds}=0.1$ V, the ON-OFF current ratio predicted by our model is in the range $\sim10-10^8$ for $T_{ox}$ between 100 and 2 nm (see Appendix B). This figure of merit, along with some key quantities such as $I_{on}$, $I_{off}$ as a function of $D_{it}$ are shown in Fig. \ref{Fig3} in order to evaluate the effect of the FLP. Again, the possible existence of FLP makes difficult an appropiate switching.\\ 
Clearly $I_{on}$ is weakly dependent on $D_{it}$ for all values of $T_{ox}$, while $I_{off}$ has a strong dependence on it, especially for smaller values of $T_{ox}$, resulting in larger values of $I_{on}/I_{off}$. For instance, the device exhibits $I_{on}/I_{off}\sim10^4$ for $T_{ox}\sim10$ nm in the Schottky limit ($D_{it}=0$), but this high value can be even gotten assuming a partial FLP with $D_{it}\sim10^{13}$ eV$^{-1}$cm$^{-2}$ at smaller $T_{ox}$ of 2 nm. 

Another interesting prediction of our model, displayed in Fig. \ref{panel1}c, is a shifting of the threshold gate voltage ($V_{th}$) induced by $V_{ds}$, pretty the same as in short-channel MOSFETs due to the DIBL effect \cite{Taur}. In Fig. \ref{Fig4}a we show the dependence of $V_{th}$ on $V_{ds}$ at a constant threshold current $I_{th}=10^{-8}$A. For comparison with the DIBL in conventional short-channel MOSFETs (tens of mV/V), the inset shows that  $\Delta V_{th}/\Delta V_{ds}$ in the barristor is three orders of magnitude larger. The bias dependent barrier lowering effect reduces as $T_{ox}$ is further reduced because the gate plays a more dominant role. An explicit quadratic relation between $V_{th}$ and $V_{ds}$ has been found taking advantage of the insensitivity of the SBH to $V_{ds}$ when $V_{ds}\gtrsim0$ and $T_{ox}$ is small enough. Details of its derivation are given in Appendix C. That expression reads as:

\begin{equation}
qV_{th}\approx a\left( \dfrac{qV_{ds}}{\eta}-b\right)  \left( \dfrac{qV_{ds}}{\eta}-b+\dfrac{1}{a}\right)+W_m-W_g,
\label{eq:Vth}
\end{equation}
where $a={q^2D_0}/(2C_{ox})$ and
\begin{equation}
b\approx k_BT\left[ log\left( \dfrac{I_{th}}{I_0u_0}\right)+1\right]  +\dfrac{qI_{th}R}{\eta}+W_s-W_g+q\phi_a
\label{eq:b}
\end{equation}

Fig. \ref{Fig4}b shows the effect of the interface trapped charge on the subthreshold voltage for several oxide thicknesses. It turns out that Vth becomes extremely sensitive to large values of $D_{it}$ (Mott limit). 
Next, we deal with the effect of the oxide thickness scaling on some figures of merit.  Fig. \ref{panel2}a shows the ouput characteristics of a barristor having $T_ {ox}=10$ nm and $D_{it}=0$, which can be compared with the case $T_{ox}=100$ nm shown in Fig. \ref{panel1}b. In the inset, we have plotted the SBH as a function of $V_g$. From it, clearly the $V_{ds}$ control over the SBH decreases when $T_{ox}$ is smaller. That is due to the strong gate control resulting in a distribution of charge, mostly, between gate metal electrode and graphene, therefore the charge in the semiconductor is small and the drain can hardly modulate it.

In Fig. \ref{panel2}b we show the average subthreshold slope at $V_{ds}=0.1$ V as a function of $T_{ox}$ with and without FLP. The presence of FLP degrades the subthreshold swing (SS), being this effect more important at large $T_{ox}$ and low permitivities. For instance, using a high-k as gate insulator (HfO$_2$) in combination with small $T_{ox}$ results in SS much closer to 60 mV/dec. Finally, Fig. \ref{panel2}c shows the value of $V_{g,off}$, as a function of $T_{ox}$, needed to keep a constant value $I_{on}/I_{off}=5\times10^4$ , again at $V_{g,on}=0.1$ V, and $V_{ds}=0.1$ V. For instance, selecting $V_{g,off}$ between 2V - 3V  an ON-OFF current ratio of $\sim10^4$ is feasible for $T_{ox}\lesssim10$ nm, in the situation of no FLP.

\section{Conclusions}
In conclusion, we have theoretically studied the electrostatics and current-voltage characteristics of the barristor device considering effects of FLP arising by possible presence of surface states, similarly to the metal-semiconductor junction. Our study suggests that the barristor is a feasible graphene logic device achieving high enough ON/OFF current ratio. When FLP dominates the barristor's electrostatics, then the gate electrode cannot modulate the SBH any more and rectification could be totally lost. On the other hand, our model has revealed that the barristor exhibits changes of the threshold voltage induced by the drain-source voltage, similarly to the Drain Induced Barrier Lowering in short channel MOSFETs. It turns out that the barristor has to be biased at low $V_{ds}$ to get a sufficient ON-OFF current ratio.  As a final note, here we have investigated the impact that a non-ideal interface might have in the barristor operation, and we have pointed out the role of oxide thickness scaling could have to get appropiate digital performance.\\

\appendices
\section{Solution of the equations.}
The non-linear system of equations 1-3, which involve both the electrostatics and the current of the device, can be understood as a system of three coupled equations where the ouput variables are $\Delta E_F, \phi_s$ and $V$ and the input parameters are $V_{ds}, V_g$, and the geometrical and electrical parameters listed the Table I. Considering that $V_{ox}=V_{ox}(\Delta E_F)$ from Eq. 2b, $\phi_b=\phi_b(\phi_s,V)$ from Eq. 2d and the definitions of the charges, we can express Eq. 2a as follows:

\begin{equation}
V_{ox}(\Delta E_F)C_{ox}+Q_g(\Delta E_F)+Q_s(\phi_s)+Q_{ss}(\phi_s,V)=0.
\label{eq:E1}
\end{equation} 

By using the definition of the voltage drop $\Delta=-\left( Q_s+Q_{ss}\right)/C_d $ across the interface layer, Eq. 2c reads as:

\begin{equation}
W_{g}+\Delta E_{F}-\dfrac{q}{C_d}\left( Q_s(\phi_s)+Q_{ss}(\phi_s,V)\right)=W_{s}-q\phi_s-qV.
\label{eq:E2}
\end{equation} 

Also, Eqs. 1 and 3 can be combined to get:

\begin{equation}
I_0R\left( \dfrac{\phi_b(\phi_s,V)}{v_t}+1\right)e^{-\phi_b(\phi_s,V)/v_t}\left[e^{V/\eta v_t}-1 \right]=V_{ds}-V.
\label{eq:E3}  
\end{equation}

In summary, Eqs. \ref{eq:E1}-\ref{eq:E3} can be rewritten and solved as a set of three non-linear coupled equations with ouput variables $\Delta E_F, \phi_s$ and $V$, namely:

\begin{subequations}
\label{eq:coupled}
\begin{align}
		 &F_1(\Delta E_F, \phi_s, V; V_{ds}, V_g,...)=0,\\
         &F_2(\Delta E_F, \phi_s, V; V_{ds}, V_g,...)=0,\\
         &F_3(\phi_s, V; V_{ds}, V_g,...)=0.
\end{align}        
\end{subequations}

\section{Additional simulations for the barristor}

\begin{figure}[h]
\centering
\includegraphics[width=90mm, height=27mm]{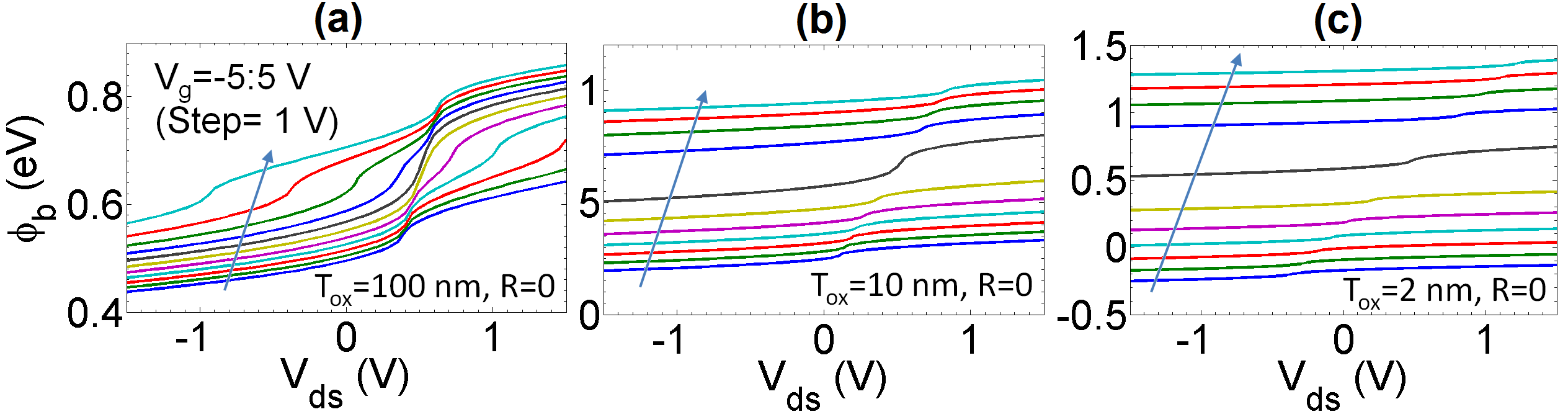}
\caption{Schottky barrier height in the graphene-semiconductor junction of the barristor as a function of $V_{ds}$ for (a) $T_{ox}=100$ nm, (b) $T_{ox}=10$ nm and (c) $T_{ox}=2$ nm. In all these cases we have assumed $D_{it}=0$ and $R=0$. }
\label{S4}
\end{figure}

\begin{figure}[h]
\centering
\includegraphics[width=90mm, height=27mm]{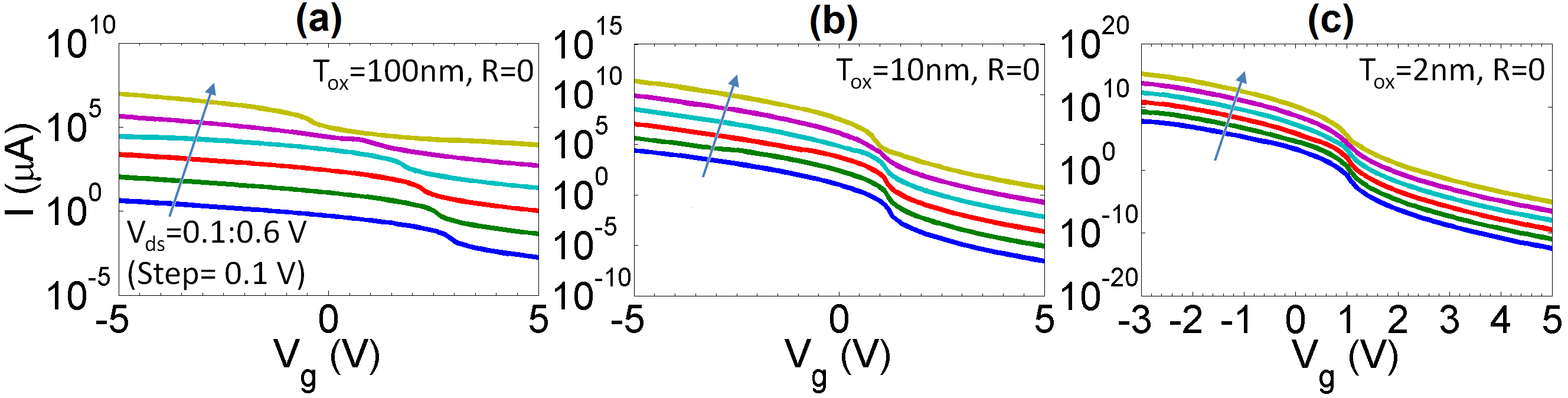}
\caption{Transfer characteristics of the barristor for several $V_{ds}$ for (a) $T_{ox}=100$ nm, (b) $T_{ox}=10$ nm and (c) $T_{ox}=2$ nm. In all these cases we have assumed $D_{it}=0$ and $R=0$.}
\label{S5}
\end{figure}

\begin{figure}[h]
\centering
\includegraphics[width=90mm, height=27mm]{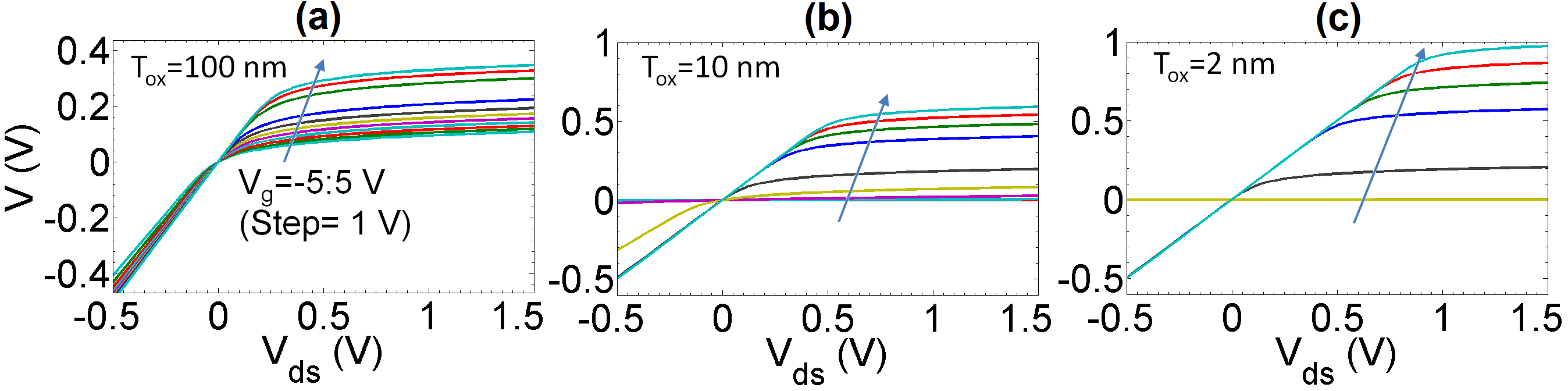}
\caption{Voltage drop across the Schottky junction as a function of $V_{ds}$ for (a) $T_{ox}=100$ nm, (b) $T_{ox}=10$ nm and (c) $T_{ox}=2$ nm. In all these cases we have assumed $D_{it}=0$ and $R=250$k$\Omega$.}
\label{S1}
\end{figure}

\begin{figure}[h]
\centering
\includegraphics[width=90mm, height=27mm]{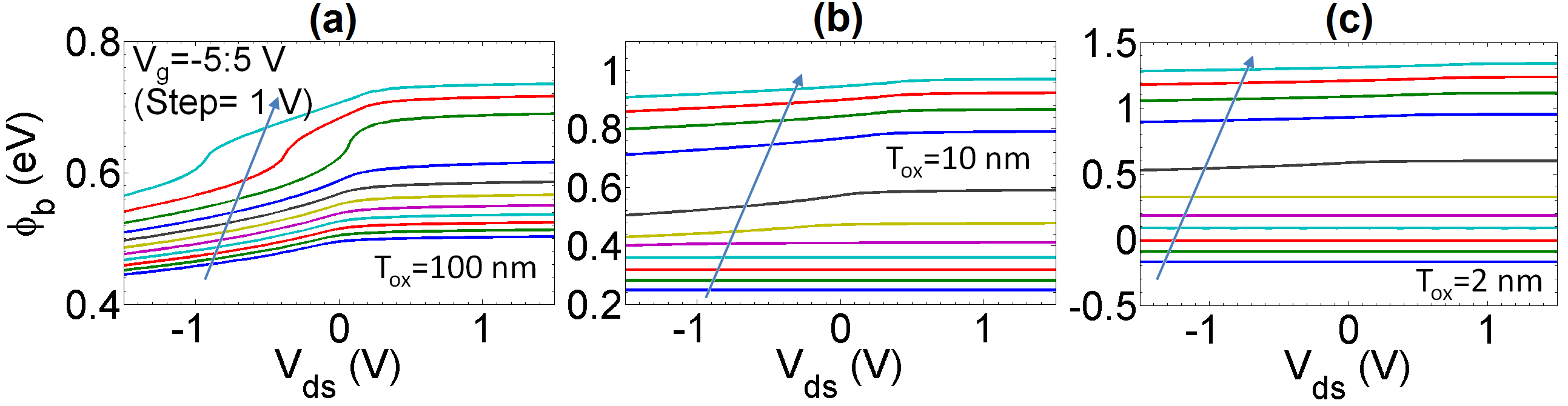}
\caption{Schottky barrier height in the graphene-semiconductor junction of the barristor as a function of $V_{ds}$ for (a) $T_{ox}=100$ nm, (b) $T_{ox}=10$ nm and (c) $T_{ox}=2$ nm. In all these cases we have assumed $D_{it}=0$ and $R=250$k$\Omega$.}
\label{S2}
\end{figure}

\begin{figure}[h]
\centering
\includegraphics[width=90mm, height=27mm]{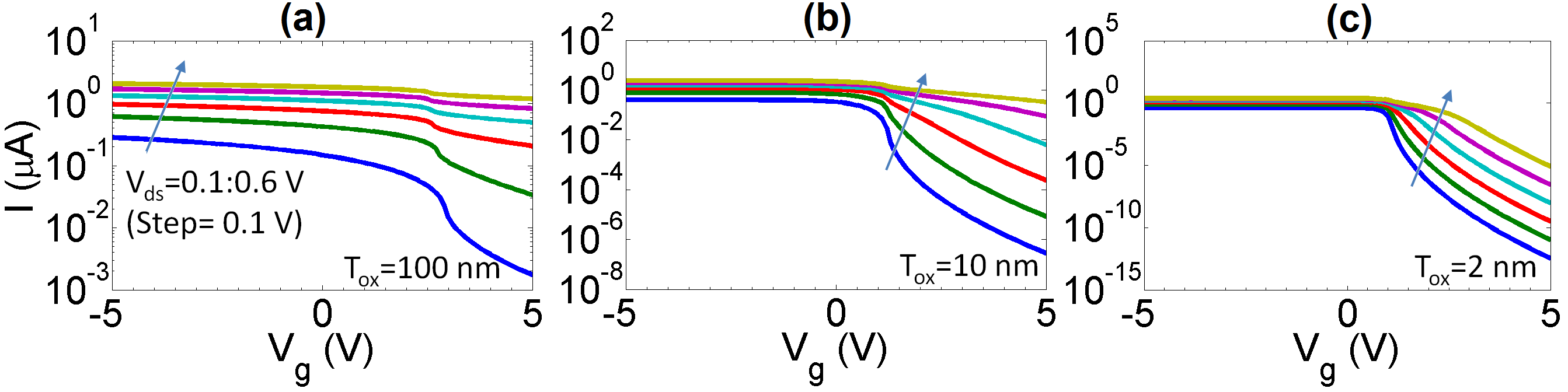}
\caption{Transfer characteristics of the barristor for several drain voltages for (a) $T_{ox}=100$ nm, (b) $T_{ox}=10$ nm and (c) $T_{ox}=2$ nm. In all these cases we have assumed $D_{it}=0$ and $R=250$k$\Omega$.}
\label{S3}
\end{figure}

In this section we show additional simulations in order to get a better understanding of the barristor's properties without FLP for several oxide thickness. We have considered both $R=0$  and $R=250$ k$\Omega$ cases in figures \ref{S4}-\ref{S5} and figures \ref{S1}-\ref{S3}, respectively. The rest of parameters are from Table I.

\section{Barristor's threshold voltage and its dependence on the drain voltage (no FLP)}

In order to obtain the barristor's threshold voltage $V_{th}$ as a function of the drain-source voltage $V_{ds}$, we start from the equation (1) of the main text, which can be rewritten as:

\begin{equation}
I=I_0\left( \dfrac{\phi_b}{v_t}+1\right)e^{-\phi_b/v_t}\left[e^{(V_{ds}-IR)/\eta v_t}-1 \right].
\label{eq:S1}
\end{equation} 
To determine $V_{th}$, let us assume the barristor biased in the off state (with $V_g>0$) and $V_{ds}\geq 0.1$ V. Under these conditions, we can safely assume $V_{ds}-IR>>3\eta v_t$ and $\phi_b>>3v_t$. Let us define now $V_{th}$ as the gate voltage needed to deliver a current $I_{th}=10^{-8}$ A.  So, Eq. \ref{eq:S1} can be approximated as:

\begin{equation}
I_{th}=I_0\dfrac{\phi_b}{v_t}e^{-\phi_b/v_t}e^{(V_{ds}-I_{th}R)/\eta v_t} .
\label{eq:S2}
\end{equation} 
After some algebra we get

\begin{equation}
V_{ds}=\eta v_tlog\left( \dfrac{I_{th}}{I_0}\right)+I_{th}R-\eta v_tlog\left( \dfrac{\phi_b}{v_t}\right)+\eta\phi_b.
\label{eq:S3}
\end{equation} 

Then we use a first order Taylor series expansion of the logarithm function, so we can write $log(u)\approx log(u_0)+u/u_0-1$ for $u$ around $u_0$, being $u_0>>1$. Using that result, we can find an expression for the SBH as an explicit function of $V_{ds}$:

\begin{equation}
\phi_b=\dfrac{1}{\left( 1-\dfrac{1}{u_0}\right)}\left\lbrace \frac{V_{ds}}{\eta}-v_t\left[ log\left( \dfrac{I_{th}}{u_0I_0}\right)+1 \right]  -\dfrac{I_{th}R}{\eta}\right\rbrace,
\label{eq:S4}
\end{equation}
where we have assumed $u_0=\phi/v_t$ is within the range 20-100 (See Figs. \ref{S4} and \ref{S2}). The expression of Eq. \ref{eq:S4} holds for any $T_{ox}$. If we further assume $T_{ox}\lesssim10$ nm, then the gate totally controls the electrostatics and the charge is distributed between the metal gate and the graphene, i. e. $Q_s\thicksim0$ and $Q_m+Q_g\thickapprox0$. Then, by combining Eqs. 2a-2b from the main text, we obtain:

\begin{equation}
\dfrac{q^2}{C_{ox}\pi\hbar^2v_f^2}\Delta E_F\vert\Delta E_F\vert+\Delta E_F+W_g-W_m+qV_g=0.
\label{eq:S5}
\end{equation} 

The solution of Eq. \ref{eq:S5} can be expressed as:

\begin{equation}
\Delta E_F=sign(\omega)\dfrac{1-\sqrt{1+4a\vert\omega\vert}}{2a},
\label{eq:S6}
\end{equation} 
where $a={q^2}/(C_{ox}\pi\hbar^2v_f^2)={q^2}D_0/(2C_{ox})$ and $\omega=W_g-W_m+qV_g$. Now, by combining Eqs. 2c-2d, an expression of the SBH as a function of the threshold voltage can be obtained:

\begin{equation}
q\phi_b=W_s+q\phi_a-W_g-sign(\omega_{th})\dfrac{1-\sqrt{1+4a\vert\omega_{th}\vert}}{2a},
\label{eq:S7}
\end{equation}
where $\omega_{th}=W_g-W_m+qV_{th}$. Finally, by replacing Eq. \ref{eq:S4} into Eq. \ref{eq:S7} and after some manipulation, an explicit relation $V_{th}=V_{th}(V_{ds})$, valid for small $T_{ox}$, is obtained:

\begin{equation}
qV_{th}=a\left[ \dfrac{qV_{ds}}{\eta*}-b\right]  \left[ \dfrac{qV_{ds}}{\eta*}-b+\dfrac{1}{a}\right]+W_m-W_g,
\label{eq:VthA}
\end{equation}  
where $\eta*=\eta( 1-{1}/{u_0}) $ and

\begin{equation}
b= \dfrac{k_BT}{\eta*}\left[  log\left( \dfrac{I_{th}}{I_0u_0}\right)+1\right]  +\dfrac{qI_{th}R}{\eta*}+W_s-W_g+q\phi_a.
\label{eq:bA}
\end{equation}

\section{benchmarking against experimental data}
In this Section we benchmark our model with two experiments reported in the literature, namely: a graphene-Si barristor working in the Schottky limit \cite{Samsung}, and a graphene-Gase barristor working in the Mott limit \cite{Kimour}.

\begin{figure}[h]
\centering
\includegraphics[scale=0.16]{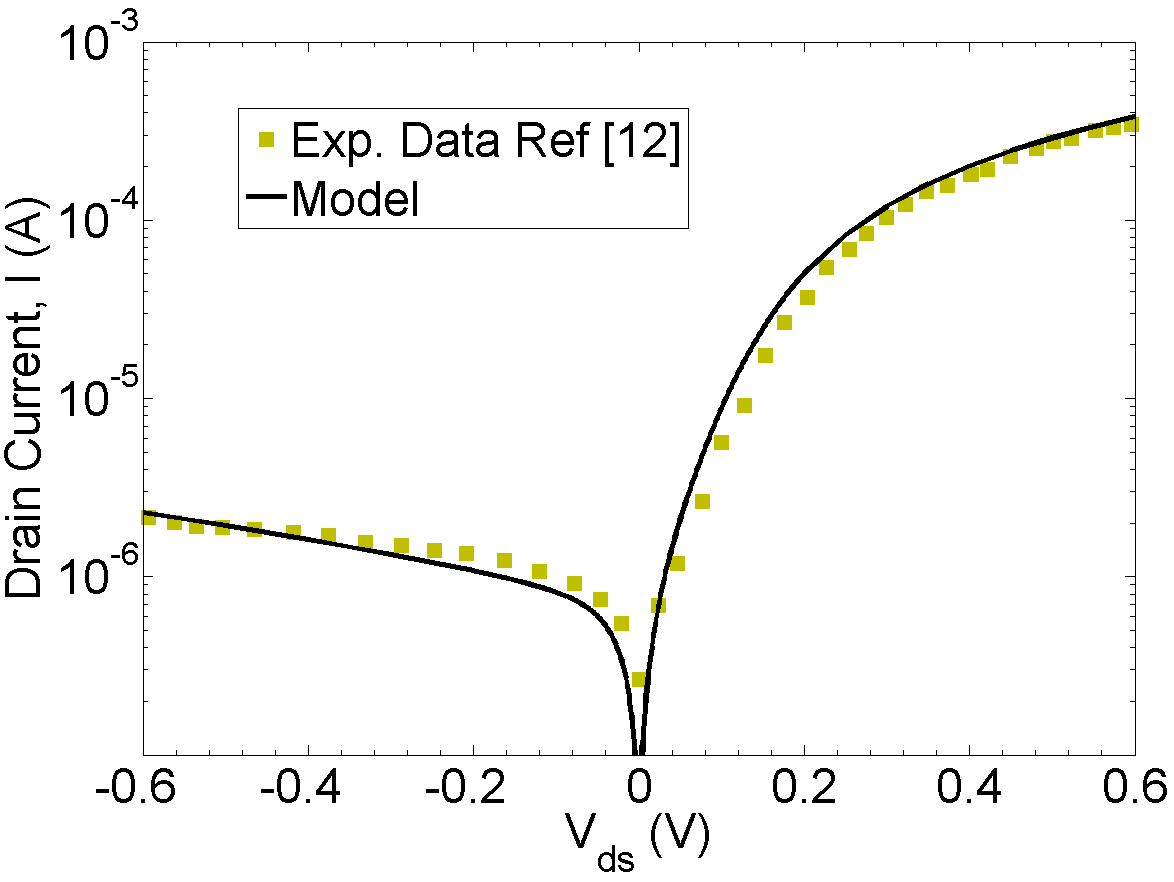}
\caption{Logarithmic I-V characteristic of a Graphene-Si barristor at $V_g=0$. Symbols: Experimental measurements from  Ref. \cite{Samsung} and solid line: results from our model in this work. To capture the trends given by the experimental data the device has been assumed to operate close to the Schottky limit, with $D_{it}=0$, $q\phi_0=0.4$ eV, $\eta=1.1$, $A=30\times10^{-5}$ cm$^2$ and $R=100$ $\Omega$.}
\label{ultima1}
\end{figure}

\begin{figure}[h]
\centering
\includegraphics[scale=0.16]{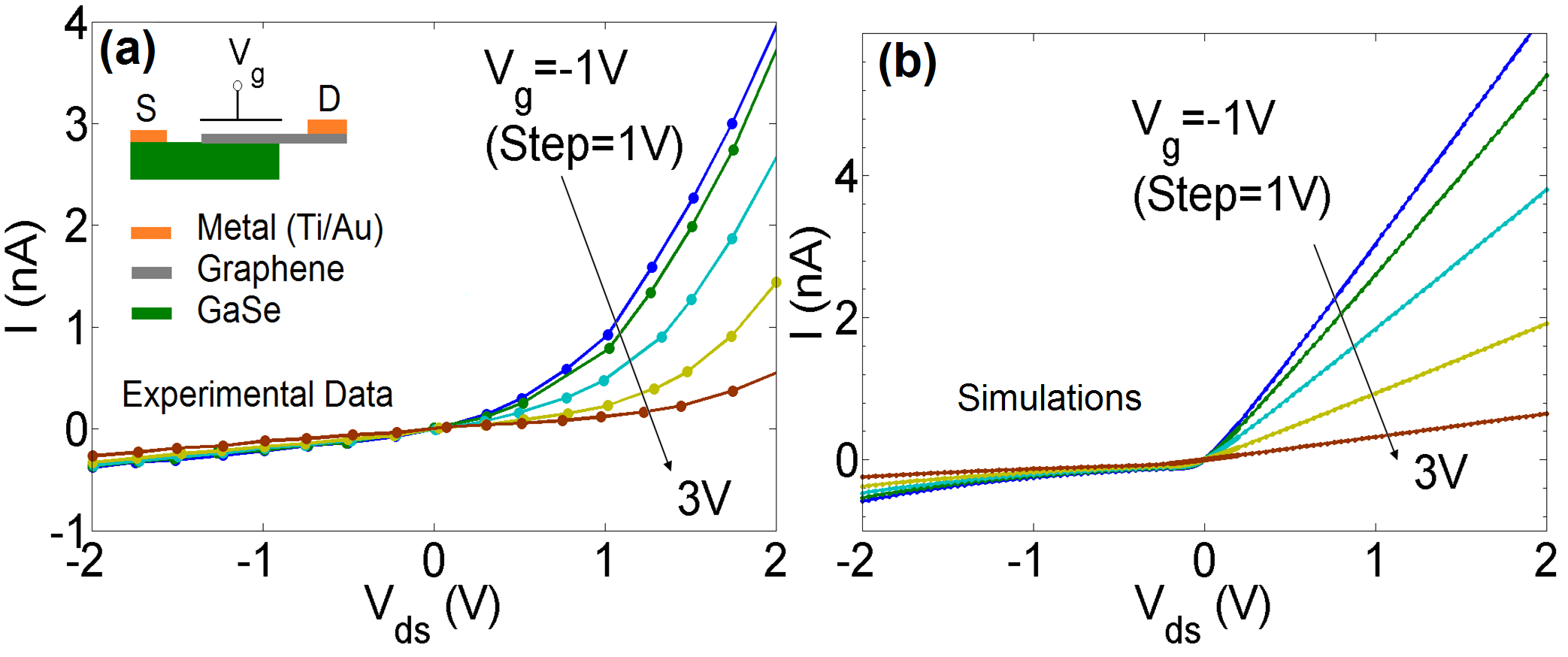}
\caption{Linear I-V characteristics	of a Graphene-GaSe barristor with $Al_2O_3$ as insulator ($T_{ox}=40$ nm). (a) Experimental measurements from Ref. \cite{Kimour} and (b) results from our model in this work. To capture the trends given by the experimental data the device has been assumed to operate close to the Mott limit, with $D_{it}=2\times 10^{14}$ eV$^{-1}$cm$^{-2}$. Other assumed parameters are: $q\phi_0=1$ eV, $\eta=1.025$ and $R$ goes between $0.3-3$ G$\Omega$.}
\label{ultima2}
\end{figure}

\section*{Acknowledgments}

This project has received funding from the European Union's Horizon 2020 research and innovation programme under grant agreement No 696656, the Department d'Universitats, Recerca i Societat de la Informaci\'o of the Generalitat de Catalunya under contract 2014 SGR 384 and the Ministerio de Econom\'ia y Competitividad of Spain under grants TEC2012-31330 and TEC2015-67462-C2-1-R (MINECO/FEDER).


\begin{thebibliography}{1}
\footnotesize


\bibitem{Tongay} S. Tongay, M. Lemaitre, X. Miao, B. Gila, B. R. Appleton, and A. F. Hebard, ``Rectification at graphene-semiconductor interfaces: Zero-gap semiconductor-based diodes," \textit{Phys. Rev. X}, vol. 2, p. 011002, Jan. 2012.

\bibitem{Parui} S. Parui, R. Ruiter, P. J. Zomer, M.Wojtaszek, B. J. vanWees, and T. Banerjeeb, ``Temperature dependent transport characteristics of graphene/n-Si diodes," \textit{J. Appl. Phys.}, vol. 116, p. 244505, 2014.

\bibitem{Yim} C. Yim, N. McEvoy, and G. S. Duesberg,``Characterization of graphene-silicon Schottky barrier diodes using impedance spectroscopy,"\textit{ Appl.Phys. Lett}, vol. 103, 193106, 2013.

\bibitem{Chen} C.-C. Chen, M. Aykol, C.-C. Chang, A. F. J. Levi, and S. B. Cronin, ``Graphene-Silicon Schottky Diodes," \textit{Nano Lett.}, vol. 11, pp. 1863-1867, April 2011.

\bibitem{Sinha} D. Sinha and J. U. Lee, ``Ideal Graphene / Silicon Schottky Junction Diodes,"\textit{ Nano Lett.}, vol. 14, pp. 4660-4664, July 2014.

\bibitem{Kopens} F.H.L. Koppens, T. Mueller, Ph. Avouris, A.C. Ferrari, M.S. Vitiello, M. Polini, ``Photodetectors based on graphene, other two-dimensional materials and hybrid systems," \textit{Nat. Nanotechnol.}, vol. 9, pp. 780-793, oct. 2014.

\bibitem{Lis}X. S. Li, Y. W. Zhu, W. W. Cai, M. Borysiak, B. Y. Han, D. Chen, R. D. Piner, L. Colombo, R. S. Ruoff, ``Transfer of Large-Area Graphene Films for High-Performance Transparent Conductive Electrodes," \textit{Nano Lett.}, vol. 9, pp. 4359-4363, Oct. 2009. 

\bibitem{Li} X. Li, H. Zhu, K. Wang, A. Cao, J. Wei, C. Li, Y. Jia, Z. Li, X. Li, and D. Wu, ``Graphene-On-Silicon Schottky Junction Solar Cells," \textit{Adv. Mater.}, vol. 22, pp. 2743-2748, July 2010.

\bibitem{Lancellotti} L. Lancellotti, T. Polichetti, F. Ricciardella, O. Tari, S. Gnanapragasam, S. Daliento, and G. Di Francia, ``Graphene applications in Schottky barrier solar cells," \textit{Thin Solid Films}, vol. 522, pp. 390-394, Nov. 2012.

\bibitem{Miao} X. Miao, S. Tongay, M. K. Petterson, K. Berke, A. G. Rinzler, B. R. Appleton, and A. F. Hebard, ``High Efficiency Graphene Solar Cells by Chemical Doping," \textit{Nano Lett.}, vol. 12, pp. 2745-2750, May 2012.

\bibitem{Kim} H.-Y. Kim, K. Lee, N. McEvoy, C. Yim, and G. S. Duesberg, ``Chemically Modulated Graphene Diodes," \textit{Nano Lett.}, vol. 13, pp. 2182-2188, April 2013.

\bibitem{Samsung} H. Yang, J. Heo, S. Park, H. J. Song, D. H. Seo, K.-E. Byun, P. Kim, I. Yoo, H.-J. Chung,K. Kim, ``Graphene barristor, a triode device with a gate-controlled Schottky barrier," \textit{Science}, vol. 336, pp. 1140-1143, Jun. 2012.


\bibitem{Kimour} W. Kim, C. Li, F. A. Chaves, D. Jim\'{e}nez, R. D. Rodriguez, J. Susoma, M. A. Fenner, H. Lipsanen, J. Riikonen, (2015, Dec.), ``Tunable Graphene-GaSe Dual Heterojunction Device," \textit{Adv. Mat.}, Available:  DOI: 10.1002/adma.201504514 (2015).

\bibitem{Dang} X. Dang, H Dong, L. Wang, Y. Zhao, Z. Guo, T. Hou, Y. Li, and S.-T. Lee, ``Semiconducting Graphene on Silicon from First-Principles  calculations," \textit{ACS Nano}, vol. 9, pp. 8562-8568, July 2015 

\bibitem{Gomila} G. Gomila and J. M. Rubí, ``Relation for the nonequilibrium population of the interface states: Effects on the bias dependence of the ideality factor," \textit{J. Appl. Phys.}, vol. 81, pp. 2674, March 1997.


\bibitem{Castro} A.H. Castro Neto, F. Guinea, N.M.R. Peres, K.S. Novoselov, and A.K. Geim, ``The electronic properties of graphene," \textit{Rev. Mod. Phys.}, vol. 81, pp. 109-162, Jan. 2009.

\bibitem{Sze} S. M. Sze, K. N. Kwok, Physics of Semiconductor Devices, \textit{John Wiley and Sons}, 2006.


\bibitem{Taur} Y. Taur, T. H. Ning, Fundamentals of Modern VLSI Devices, \textit{Cambridge University Press}, May 2013

\end{thebibliography}
\end{document}